\documentstyle[12pt,world_sci]{article}

\def\PRL#1{{\em Phys.~Rev.~Lett.~}{\bf #1}}
\def\PRD#1{{\em Phys.~Rev.~}{\bf D#1}}
\def\NPB#1{{\em Nucl.~Phys.~}{\bf B#1}}
\def\PLB#1{{\em Phys.~Lett.~}{\bf B#1}}

\def\ImPi{{\rm Im}\Pi}

\def\ImDelta{{\rm Im}\Delta}

\def\drho{\delta\rho}
\def\drhot{\delta\rho^{\rm T}}
\def\drhol{\delta\rho^{\rm L}}
\def\dRT{\delta R^{\rm T}}
\def\dRL{\delta R^{\rm L}}

\def\beq{\begin{equation}}
\def\eeq{\end{equation}}
\def\beqa{\begin{eqnarray}}
\def\eeqa{\end{eqnarray}}

\begin{document}

\begin{flushleft}
Presented at Beyond the Standard Model IV \hfill  FERMILAB--CONF--95/084--T \\
Lake Tahoe, California, Dec. 13--18, 1994 \hfill  EFI 95--19 \\
Proceedings to be published by World Scientific \hfill April 1995 \\
\end{flushleft}

\title{{\bf THE DIFFICULTIES INVOLVED IN CALCULATING $\drho$.}}

\author{TATSU TAKEUCHI\thanks{Presenting author} \\
       {\em Fermi National Accelerator Laboratory \\
            P.O. Box 500, Batavia, IL 60510
       } \\
       \vspace{0.3cm}
       and \\
       \vspace*{0.3cm}
       AARON K. GRANT and MIHIR P. WORAH \\
       {\em Enrico Fermi Institute and Department of Physics,
            University of Chicago \\
            5640 S. Ellis Avenue, Chicago, IL 60637
       }
       }

\maketitle

\setlength{\baselineskip}{2.6ex}

\begin{center}
\parbox{13.0cm}
{\begin{center} ABSTRACT \end{center}
{\small \hspace*{0.3cm}
We discuss the difficulties that arise when one tries to
calculate $\drho$ using dispersion relations.
}}
\end{center}

\medskip
\section{Introduction}

Recently, several authors have attempted to make use of
dispersion relations (DR's)
to calculate the contribution of non--relativistic (NR)
bound states and threshold enhancements
to the $\rho$ parameter. \cite{HUNG,KS}
The idea was to write the $\rho$ parameter as a dispersion integral
over the imaginary parts of the vacuum polarization functions,
and then to
use the Schr\"odinger equation to calculate the
contributions of NR intermediate states to these
spectral functions.

In Ref.~1, the contribution of possible 4th generation
fermion--antifermion bound states formed from Higgs exchange
was calculated, while Ref.~2 calculated the contribution
of the $t\bar{t}$ threshold enhancement due to QCD binding effects.

In this short note,
we would like to point out that these calculations suffer from
a fundamental flaw:  the sign of the contribution of
NR states to the $\rho$ parameter is actually
indeterminable.  This is due to the fact that it is possible to
write two different DR's for $\drho$ and the contribution of
NR states changes sign depending on which DR is used.

In the following, we will give a brief discussion on how this
comes about.

%\medskip
\section{The Dispersion Relations}

Following Ref.~2 we introduce the following notation
for current--current correlation functions:
\beqa
\Pi_{\mu\nu}(q)
& = & -i\int d^4x e^{iq\cdot x}
      \langle 0 | T^* \left[ J_\mu(x) J_\nu(0)
                      \right]
      | 0 \rangle  \cr
& = & g_{\mu\nu} \Pi(s) + q_\mu q_\nu \lambda(s)
      \phantom{\left( \frac{q_\mu q_\nu}{q^2}
               \right)} \cr
& = & \left( g_{\mu\nu} - \frac{q_\mu q_\nu}{q^2}
      \right) \Pi(s)
    + \left( \frac{q_\mu q_\nu}{q^2}
      \right) \Delta(s)
\eeqa
where $s=q^2$. Note that
\beq
\Delta(s) = \Pi(s) + s\lambda(s).
\eeq
Therefore,
\beq
\Delta(0) = \Pi(0)
\eeq
unless $\lambda(s)$ has a pole at $s=0$.
Now the shift of the $\rho$ parameter away from its
tree level value of $\rho=1$ is usually expressed as
the difference between the {\it transverse} parts
of the charged and neutral isospin current
correlators:
\beq
\drho = \frac{1}{v^2}\left[ \Pi_{+-}(0) - \Pi_{33}(0)
                     \right],
\label{TRANS}
\eeq
where $v\approx 246$GeV is the Higgs VEV.
However, if we are considering the contribution of the
3rd or 4th generation fermions to $\drho$ as in
Refs.~1 and 2, one can also
express $\drho$ as a difference between the
{\it longitudinal} parts since the $\lambda$--functions
are free of poles at $s=0$.  Hence,
\beq
\drho = \frac{1}{v^2}\left[ \Delta_{+-}(0) - \Delta_{33}(0)
                     \right].
\label{LONG}
\eeq
Applying Cauchy's theorem to Eq.~\ref{TRANS},
we find the DR
\beq
\drho = \drhot(\Lambda^2) + \dRT(\Lambda^2),
\label{TDR}
\eeq
where
\beqa
\drhot(\Lambda^2)
& \equiv & \frac{1}{v^2}
           \left[ \frac{1}{\pi}
                  \int^{\Lambda^2} \frac{ds}{s}
                  \left\{ \ImPi_{+-}(s)
                        - \ImPi_{33}(s)
                  \right\}
           \right],  \cr
\dRT(\Lambda^2)
& \equiv & \frac{1}{v^2}
           \left[ \frac{1}{2\pi i}
                  \oint_{|s|=\Lambda^2}
                  \left\{ \Pi_{+-}(s)
                        - \Pi_{33}(s)
                  \right\}
           \right].
\eeqa
This was the DR used in Ref.~1.
Alternatively, we can apply Cauchy's theorem to
Eq.~\ref{LONG} and obtain
\beq
\drho = \drhol(\Lambda^2) + \dRL(\Lambda^2),
\label{LDR}
\eeq
where
\beqa
\drhol(\Lambda^2)
& \equiv & \frac{1}{v^2}
           \left[ \frac{1}{\pi}
                  \int^{\Lambda^2} \frac{ds}{s}
                  \left\{ \ImDelta_{+-}(s)
                        - \ImDelta_{33}(s)
                  \right\}
           \right],  \cr
\dRL(\Lambda^2)
& \equiv & \frac{1}{v^2}
           \left[ \frac{1}{2\pi i}
                  \oint_{|s|=\Lambda^2}
                  \left\{ \Delta_{+-}(s)
                        - \Delta_{33}(s)
                  \right\}
           \right].
\eeqa
This was the DR used in Ref.~2.
We keep the radius of the integration contour $\Lambda^2$ finite
in our expressions since $\dRT(\Lambda^2)$ and $\dRL(\Lambda^2)$
may not necessarily converge to zero as $\Lambda^2 \rightarrow \infty$.
\cite{TGW}

%\newpage
\medskip
\section{Contribution of Non--Relativistic States}

Now let us consider the contribution of NR states to
$\drhot(\Lambda^2)$ and $\drhol(\Lambda^2)$.   In the NR
limit, the imaginary parts of $\Pi(s)$ and $\Delta(s)$
can be obtained by solving for the Green's function of
the Schr\"odinger equation. \cite{SP}
In particular, the bound state contributions are proportional
to the value of the NR wavefunction evaluated at the
origin squared.  Spin 1 states contribute to $-\ImPi(s)$ while
spin 0 states contribute to $\ImDelta(s)$.  Since the NR
potential has no spin dependent term, the spin 1 vector and
spin 0 pseudoscalar
states will have the exact same wavefunction.  Therefore,
the contribution of NR states to the spectral functions satisfies
\beq
-\ImPi_{+-}^{\rm NR}(s) = \ImDelta_{+-}^{\rm NR}(s),\qquad
-\ImPi_{33}^{\rm NR}(s) = \ImDelta_{33}^{\rm NR}(s),\qquad
\eeq
which implies
\beq
\drhot_{\rm NR} = -\drhol_{\rm NR}.
\eeq
This shows that the sign of the contribution of NR states to
the $\rho$ parameter depends on the DR used.   Since both
Eqs.~\ref{TDR} and \ref{LDR} must give the same
result, the contribution of the relativistic states
away from the threshold and the possible contribution from the
integral around the circle at $|s|=\Lambda^2$ must
account for the difference.

\medskip
\section{Conclusion}

We have shown that the sign of the contribution of NR states to
$\drho$ depends on the DR used and is therefore indeterminable.
One must include the contribution from the entire
integration contour to get a meaningful result.

\vspace{0.5cm}
{\bf \noindent Acknowledgements \hfil}
\vspace{0.4cm}

This work was supported by
the United States Department of Energy under
Contract Number DE--AC02--76CH03000 (T.T.), and in part
under Grant Number DE--FG02--90ER40560 (A.K.G. and M.P.W.).

\newpage
\bibliographystyle{unsrt}

\end{document}